\documentclass[letterpaper, 10 pt, conference]{ieeeconf}  % Comment this line out if you need a4paper

\IEEEoverridecommandlockouts                              % This command is only needed if
\usepackage{enumerate}
\usepackage{amsmath}
\usepackage{amssymb}
\usepackage{epstopdf} %converte l'immagine .ps o .eps in pdf per evitare problemi in compilazione
\usepackage{graphicx}      % include this line if your document contains figures
\usepackage{booktabs}
\usepackage[english]{babel}
\newtheorem{remark}{Remark}

\newtheorem{definition}{Definition}

\newtheorem{property}{Property}
\newtheorem{thm}{Theorem}

\title{\LARGE \bf
	Echo State Networks: analysis, training and predictive control}

\author{Luca Bugliari Armenio$^{1}$, Enrico Terzi$^{2}$, Marcello Farina$^{2}$, and Riccardo Scattolini$^{2}$
	\thanks{$^{1}$ The author is with the Dipartimento di Elettronica, Informazione e Bioingegneria, Politecnico di Milano, Via Ponzio 34/5, 20133, Milano, Italy. E-mail: {\tt\small luca.bugliari@mail.polimi.it}\newline
$^{2}$ The authors are with the Dipartimento di Elettronica, Informazione e Bioingegneria, Politecnico di Milano, Via Ponzio 34/5, 20133, Milano, Italy. E-mail: {\tt\small name.surname@polimi.it}}}

\begin{document}

\maketitle
\thispagestyle{empty}
\pagestyle{empty}

\begin{abstract}                % Abstract of not more than 250 words.
The goal of this paper is to investigate the theoretical properties, the training algorithm, and the predictive control applications of Echo State Networks (ESNs), a particular kind of Recurrent Neural Networks.
First, a condition guaranteeing incremetal global asymptotic stability is devised. Then, a modified training algorithm allowing for dimensionality reduction of ESNs is presented.
Eventually, a model predictive controller is designed to solve the tracking problem, relying on ESNs as the model of the system. Numerical results concerning the predictive control of a nonlinear process for $pH$ neutralization confirm the effectiveness of the proposed algorithms for the identification, dimensionality reduction, and the control design for ESNs.
\end{abstract}
%===============================================================================

\section{Introduction}
In the last decades, Recurrent Neural Networks (RNN), also called \textit{Dynamical Reservoirs}, have become very attractive for their potentially wide range of applications. RNNs were developed in the 1980s, and in 1993 it was possible to solve a deep learning task requiring more than 1000 subsequent layers in a RNN \cite{ab}. Afterwards, their use spread in a lot of different areas such as linguistic \cite{schmidhuber2002learning},\cite{gers2001lstm}, musical \cite{gers2002learning}, technological \cite{mayer2008system},\cite{graves2009offline}, social \cite{baccouche2011sequential},\cite{graves2013speech}, economic \cite{tax2017predictive} and biological \cite{thireou2007bidirectional}. 

Among RNNs, Echo State Networks (ESNs) stand out for their versatility and ease of use. The main advantage of the ESNs with respect to other RNNs lies in the training algorithm. In fact, while in general the training algorithms of RNNs are based on iterative solutions to non-linear problems \cite{hagan1994training}, e.g. the Back-propagation algorithm, the training of ESNs reduces to the solution to a linear regression problem. 
Their mathematical models in state-space form are composed of a non-linear state equation but a linear output one. The reservoir of ESN is the recurrent layer formed by a large number of sparsely interconnected units with non-trainable weights, that are chosen randomly. The ESN training procedure, indeed, is a simple adjustment of output weights to fit input/output data with the state trajectory of the net starting from a random initialization. 
ESNs enjoy, under mild conditions, the so-called \textit{Echo state property} (see \cite{ESN_Jaeger}), that ensures that the effect of the initial condition vanishes after a finite transient. ESNs have been exploited so far with notable results in speech recognition \cite{verstraeten2006reservoir}, time-series prediction tasks \cite{jaeger2004harnessing}, reinforcement learning \cite{szita2006reinforcement}, and language modelling \cite{tong2007learning}. A common feature of ESN is the large number of states, that enables the modeling of complex phenomena at the price of a significant computational effort when used in control applications. It is typical to find ESNs with thousands of states \cite{ESN_Jaeger}, which expand their descriptive capabilities but make them less suitable in cases where online optimization problems are involved, for example, when predictive control is employed. 
The use of ESNs for control purposes has been therefore rather limited so far, and only few contributions are found in control literature, among these \cite{salmen2005echo},\cite{ploger2003echo} and \cite{pan2012model}. In particular in \cite{pan2012model} the issue of nonlinearity with high dimensional states (of ESNs) is tackled by means of a Taylor expansion, to which a nonlinear term is added.

%Given that MPC schemes rely on a dynamical model of the system, and the ESNs proved to have good modeling properties, we designed a nonlinear MPC for tracking problem including ESN.	
In this work ESNs are addressed both theoretically, and through their application for control purposes, specifically in the Model Predictive Control framework \cite{camacho2013model}, that is made possible thanks to the addition of a suitable model reduction phase.
First, a sufficient condition for ESN to be \textit{Incrementally Globally Asymptotically Stable} ($\delta GAS$, \cite{Discrete-timedeltaISS}) is provided. Second, to face the dimensionality issue, we propose a modified algorithm for the training, that is based on the solution of a LASSO program \cite{tibshirani1996regression}. This allows to train a network that is non-minimal, so that its dimension can be reduced thanks to a suitable algorithm detailed in \cite{albertini1995recurrent} and based on decomposition of the system into observable/unobservable parts. Third, we consider ESN into the MPC design of a regulator solving tracking and disturbance rejection problems. Numerical examples show the effectiveness of the proposed approach for identification, dimensionality reduction, and predictive control of a SISO nonlinear plant for PH neutralization.

The paper is organized as follows: In Section \ref{Sec:notation} the notation is introduced and some properties of nonlinear systems are recalled, Section \ref{Sec:Echo} defines the model of ESNs, states the main theoretical result, and illustrates the proposed modified training algorithm. Section \ref{mpc} describes the formulation of the Model Predictive Control problem, followed by the simulation example reported in Section \ref{Sec:Simulation}. Conclusions and hints for future work are included in Section \ref{Sec:Conclusion}.

\section{Notation and basic definitions} \label{Sec:notation}
	
	Let us consider the general discrete-time system:
	\begin{equation}
	\label{DiscreteSys}
	x(k+1)=f(x(k),\omega(k))
	\end{equation}
	where $k \in \mathbb{Z}$ is the discrete time index, $x(k)$ is the state vector, $\omega(k)$ is the input vector and $f(\cdot)$ is a nonlinear function of the input and the state.
	
	Defining the sequence $\vec{\omega}=(\omega(0),\dots,\omega(k))$ for $k \geq 0$, we indicate with $x(k,x_{0},\vec{\omega})$ the solution to system \eqref{DiscreteSys} at time step $k$ starting from initial state $x_{0}$ with input sequence $\vec{\omega}$. Moreover, given a column vector $v$ with entries $v_i, i=1,\dots, n$ we denote $\|v\|_2$ as the 2-norm of $v$, $\|v\|_1=\sum_{i=1}^n|v_i|$ the 1-norm, $\|v\|_A=v^TAv$ the norm weighted by matrix $A$, and $v^T$ as the transpose of $v$. We denote with $Ker(A)=\{v\in \mathbb{R}^n:Av=0\}$ the \textit{kernel} of matrix $A$, and with $\|A\|=\sup_{v \neq 0}\frac{\|Av\|_2}{\|v\|_2}$ its norm, while $\rho(M)$ is the spectral radius of the square matrix M (i.e. maximum absolute value of the eigenvalues).
	
	In the following we also recall some notions related to nonlinear systems \cite{Discrete-timedeltaISS}, that are useful for the analysis of ESNs.
	%\begin{definition}[Null sequence \cite{NullSequence}]
	%	\label{NS}
	%	Let $X$ be one of the standard number fields $\mathbb{Q,R,C}$.
	%	Let $(x_{n})$ be a sequence in $X$ which converges to a limit of 0:
	%	\begin{equation*}
	%	\lim\limits_{n\to\infty}x_{n}=0
	%	\end{equation*}		
	%	Then $(x_{n})$ is called a null sequence.
	%\end{definition}
	\begin{definition} [$\mathcal{K}$-Function]
		A continuous function $\alpha: \mathbb{R}_{\geq0}\to\mathbb{R}_{\geq0}$ is a class $\mathcal{K}$ function if $\alpha(s)>0$ for all $s>0$, it is strictly increasing, and $\alpha(0)=0$.
	\end{definition}
	
	\begin{definition} [$\mathcal{K}_{\infty}$-Function]
		A continuous function $\alpha: \mathbb{R}_{\geq0}\to\mathbb{R}_{\geq0}$ is a class $\mathcal{K}_{\infty}$ function if it is a class $\mathcal{K}$ function and $\alpha(s)\to\infty$ for $s\to\infty$.
	\end{definition}
	\begin{definition} [$\mathcal{KL}$-Function]\label{def:KL}
		A continuous function $\beta: \mathbb{R}_{\geq0}\times\mathbb{Z}_{\geq0}\to\mathbb{R}_{\geq0}$ is a class $\mathcal{KL}$ function if $\beta(s,t)$ is a class $\mathcal{K}$ function with respect to $s$ for all $t$, it is strictly decreasing in $t$ for all $s>0$, and $\beta(s,t)\to0$ as $t\to\infty$ for all $s>0$.
	\end{definition}
	\begin{definition} [$\delta$GAS]
		A system of the form \eqref{DiscreteSys} is called \textit{incrementally globally asymptotically stable} ($\delta GAS$), if there exist a $\mathcal{KL}$ function $\beta$ such that for all $k\in \mathbb{Z}_{\geq0}$, any initial states $x_{0},x'_{0}$ and any disturbance sequence $\vec{\omega}$
		\begin{equation*}
		\|x(k,x_{0},\vec{\omega})-x(k,x'_{0},\vec{\omega})\|_2\leq\beta(\|x_{0}-x'_{0}\|_2,k)
		\end{equation*}
	\end{definition}

\section{Echo State Networks}\label{Sec:Echo}
	\subsection{Model}
	In the terminology of Neural Networks, ESNs are composed of a dynamical reservoir (also called hidden layer) in which the connections between neurons are sparse and random. 
	The ESN we consider is characterized by $n$ neurons in the reservoir, i.e. states composing the vector $x$, an input $u$ and an output $y$. The nonlinear activation function of the neurons is $\tanh(\cdot)$, as is common practice \cite{ESN_tut}. The state and output equations of the ESN read:
	\begin{alignat}{2}
	x(k+1) & = \tanh(W_xx(k)+W_{u}u(k)+W_{y}y(k)) 
	\label{net_state}\\[1.5ex]
	y(k) & = W_{out_{1}}x(k)+W_{out_{2}}u(k-1)
	\label{net_output}
	\end{alignat}
	
	where $W_{u}, W_x, W_{y}, W_{out_{1}}$ and $W_{out_{2}}$ represent the connection weight matrices. 
The form \eqref{net_state} - \eqref{net_output} is equivalent to the conventional one \cite{ESN_Jaeger}, the equivalence is shown in Appendix (Property \ref{Prop_alternativeSS}). 
Matrices $W_x,W_y, W_u$ are randomly generated, while $W_{out_1}$ and $W_{out_2}$ are not known, and must be properly identified in the training phase. For simplicity of presentation, here we consider a SISO system, i.e. $y(k)$ and $u(k)$ are scalar variables.

\subsection{Properties}	
	With reference to model \eqref{net_state}, the notion of \textit{Echo State} plays an important role in understanding the ESN behaviour. 
	It is related to the weight matrices $(W_x,W_{u},W_{y})$, and thus to the network before its training, and it takes the following definition \cite{ESN_Jaeger}.
	
	\smallskip

	\begin{property} [Echo States]
		\label{ES}
		Assume to have a network \eqref{net_state}, \eqref{net_output} with weights $(W_x,W_{u},W_{y})$ driven by an input $u(k)$ and with output $y(k)$ belonging to compact intervals $\mathcal{U}$ and $\mathcal{Y}$. \newline The network $(W_x,W_{u},W_{y})$ has \textit{echo states} with respect to $\mathcal{U}$ and $\mathcal{Y}$, if for every left-infinite input/output sequence $(u(k-1),y(k-1))$, where $k \in (-\infty,0)$, and for all state sequences $x'(k),x''(k)$ subject to the network dynamics:
		\begin{alignat*}{2}
		x'(k+1) & = \tanh(W_{u}u(k)+W_xx'(k)+W_{y}y(k))\\
		x''(k+1) & = \tanh(W_{u}u(k)+W_xx''(k)+W_{y}y(k))
		\end{alignat*}
		it holds that $x'(k)= x''(k)$ for all $k\geq 0$. \hfill $\square$
	\end{property}

\smallskip

	In other words, if the ESN has been run for a very long time, the current network state does not depend on its initial value, but only on the history of the forcing signal. Experimental results \cite{ESN_Jaeger} show empirically that a sufficient condition for the \textit{Echo State Property} is $\rho(W_x)<1$. 
In the following, $\delta GAS$ is formally guaranteed under a slightly stronger condition, i.e. $\|W_x\|<1$.

\smallskip

\begin{thm}
		\label{thm1}
		If $\|W_x\|<1$, then system \eqref{net_state} is $\delta GAS$ \hfill $\square$
\end{thm}
\smallskip
	\begin{proof}
	See the Appendix 
	\end{proof}

	\subsection{ESN Training and order reduction}
	\label{secTraining}
%	During the training of Echo State Networks we have considered the tuning of some parameters including the number of states $n$, the spectral radius $\rho(W_x)$, a feedback scaling parameter $k_y$ and various normalization factors.
	The training of ESNs consists of learning a model such that the difference between the model and the system outputs is minimized. 
	In this form, this is a Least Square (LS) problem, that allows to find the output connection matrices $W_{out_1}$ and $W_{out_2}$ (see \cite{ESN_Jaeger}). The learning algorithm is reported next.

\smallskip

\textit{Algorithm 1}
	
	\begin{enumerate}
		\item Generate randomly a sparse matrix $W_x$ with spectral radius $\rho<1$.
		\item Generate random matrices $W_u$ and $W_y$.
		\item Start from $x(0)$ arbitrary and integrate \eqref{net_state} forced by the real system data retrieved $(u,y)$.
		\item Discard the first $K_0$ points (associated to the initial conditions) and get the state trajectory $x(k)$ for $k>K_0$.
		\item Store the values of $(x(k),u(k-1))$ and of $y_{sys}(k)$ for $k>K_0$ respectively into matrices $\Phi$ and $Y_{sys}$.
		\item Solve the LS problem $\min_{W_{out}}\|Y_{sys}-\Phi W_{out}\|_2^2$ to find the output weights $W_{out}=[W_{out_1} W_{out_2}]^T$.
	\end{enumerate}
	
%	The main advantage of ESNs lies in the fact that they have a simpler training procedure since it is just a linear regression problem while the most recurrent neural networks adopt an iterative learning algorithm (e.g. backpropagation algorithm).
%	As said before the ESNs are used to model highly nonlinear systems, thus it is common to have a very large number of states since the bigger the dimension of the reservoir, the easier is to find a combination to approximate the real output. \newline
\smallskip
As widely known, however, NNs may require, in order to well reproduce the system input-output behaviour, a significant number of states, compromising the practical use of NN-based control algorithms in real-time applications. A network subject to a suitable model reduction phase may allow to reduce significantly the computational time required for the solution to the optimization program underlying the design of a MPC algorithm as described next, without meaningful losses in the modeling performances.
In this paper we propose a theoretically sound method for reducing the number of states of the network. This method relies on the notion of minimality, which has been characterized in the context of Mixed Networks (MN, see \cite{albertini1995recurrent}). Indeed, ESNs are included in this class, as shown in Appendix (Property \ref{Prop_MN}). A remarkable property of MNs is \textit{minimality}, whose definition is reported in \cite{albertini1995recurrent}. 
In a few words, a MN is \textit{minimal} if, for any initialization, any other input-output equivalent network has greater or equal dimension.
A sufficient condition for minimality is stated below \cite{albertini1995recurrent}: 
	\smallskip
	\begin{property}
		Given an ESN of the form \eqref{net_state}-\eqref{net_output}, if $W_{out_1}$ has all non-zero columns and $Ker(W_{out2})=0$, then the network is \textit{minimal}. \hfill $\square$
		\label{Prop_minimality}
	\end{property}
	
\smallskip
	
Note that, however, the solution to the LS problem stated above leads the output weights to be in general different from zero, hence the network is minimal (and not reducible) thanks to Property \ref{Prop_minimality}.

We propose a modified training algorithm that is aimed at enforcing a sparse structure to $W_{out}$ when possible, making the network non-minimal and so ``reducible''.

The modified training Algorithm consists of three steps.
\smallskip

\textit{Algorithm 2}
\begin{enumerate}
\item Perform steps 1)-6) of Algorithm 1 with the modified cost function  $$\min_{W_{out}}\|Y_{sys}-\Phi W_{out}\|_2^2+ \lambda\|W_{out}\|_1, \lambda \in \mathbb{R}^+$$ \label{LASSO}
\item Perform a dimensionality reduction of the network obtained, applying the procedure proposed in \cite{albertini1995recurrent}, pag.29. The resulting network will be \textit{minimal} and characterized by $n_0\leq n$ states\label{dim_red}
\item Perform Algorithm 1 with the reduced ESN. \label{retraining}
\end{enumerate}

Note that Step \ref{retraining}, involving the reduced ESN, is recommendable after the first training since the LASSO formulation in Step \ref{LASSO} biases the $W_{out}$ coefficients towards zero, see also \cite{friedman2001elements}. 

\section{Model Predictive Control (MPC)}
\label{mpc}
\begin{figure}[h]
		\centering
		\includegraphics[scale=0.46]{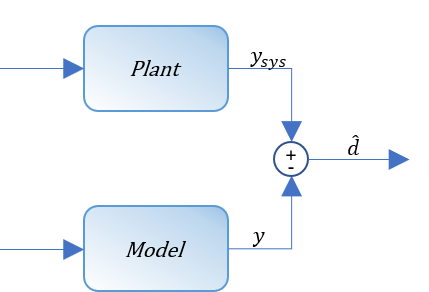}
		\caption{Calculum of the mismatch between real and estimated output}
		\label{fig: compdist}
	\end{figure}

	MPC is a model-based iterative control technique: at each sampling time $T_s$, we measure or estimate the current state of the system model (in our case the ESN), we compute the predicted output by means of \eqref{net_state} and \eqref{net_output} as function of the future control variables, and we obtain the optimal control input by minimizing a cost function subject to constraints, see \cite{maciejowski2002predictive}.
	The aim of our control strategy is to track changes of the output reference $\bar{y}$, without the need to compute the steady state values of the state and of the control input corresponding to $\bar{y}$. To this end, an integrator is inserted in cascade to the system input $u(t)$, in such a way that the actual manipulable input is now $\delta u(k)$:
	\begin{align}
	\label{c}
	u&(k)=u(k-1)+\delta u(k)
%\\[1.5ex]
%	\label{d}
%	u&(k)=v(k)
	\end{align}
	Furthermore, to avoid steady state modelling errors, a well known technique in MPC is applied, inspired by \cite{morari2012nonlinear}. This consists in considering the difference between the real system output $y_{sys}(\cdot)$ and the predicted output $y(\cdot)$ as an additional disturbance $\hat{d}(\cdot)$ (Fig. \ref{fig: compdist}).
	The estimate of the disturbance is computed each time a new optimization problem is set and the disturbance is considered constant over the entire prediction horizon $N$. Then, this estimated disturbance $\hat{d}$ is added to the future predicted outputs of the optimal control problem.
	The resulting cost function formulated for computing the optimal control input over the prediction horizon $N$ is:
	\begin{equation}
	\label{costfun}
	J(k,\delta U(k))=\sum_{i=0}^{N-1}\biggl(\|\hat{e}(k+i)\|^2_{Q}+\|\delta u(k+i)\|^2_{R}\biggr)
	\end{equation}
	where $\hat{e}(\cdot)=y_*(\cdot)-\bar{y}=y(\cdot)+\hat{d}(\cdot)-\bar{y}$ is the estimated output error, $\delta u(k)= u(k)-u(k-1)$ is the input rate of change, thus the input of the integrator, $\delta U(k)= \begin{bmatrix}\delta u(k) & \dots & \delta u(k+N-1)\end{bmatrix}$ and the matrices $Q>0$ and $R>0$ weight the estimated error and the rate of change of the input respectively. 
	The cost function $J$ is subject to the ESN dynamics \eqref{net_output} and to the constraints:
	\begin{alignat}{3}
	\label{constraint1}
	&\hat{d}(k+i)=\hat{d}(k) \qquad \qquad &i=1,\dots,N-1 \\[1.5ex]
	\label{constraint2}
	&u_{min}\leq u(k+i)\leq u_{max} \qquad \qquad &i=0,\dots,N-1
	\end{alignat}
	%plus additional constraints on the future values of $\hat{y}$ and $\delta u$. 
	After each time step $k$, the optimization problem 
	\begin{equation}
	\min_{\delta U(k)} J(k,\delta U(k)) \quad 	s.t. \quad \eqref{constraint1},\eqref{constraint2}
	\end{equation} 
	is solved, and the input $u(k)=u(k-1)+\delta u(k)$ is applied to the system according to the Receding Horizon principle, see \cite{maciejowski2002predictive}.
	
	\section{Simulation results}\label{Sec:Simulation}
	\subsection{pH neutralization process}
\begin{figure}[h]
		\centering
		\includegraphics[scale=0.5]{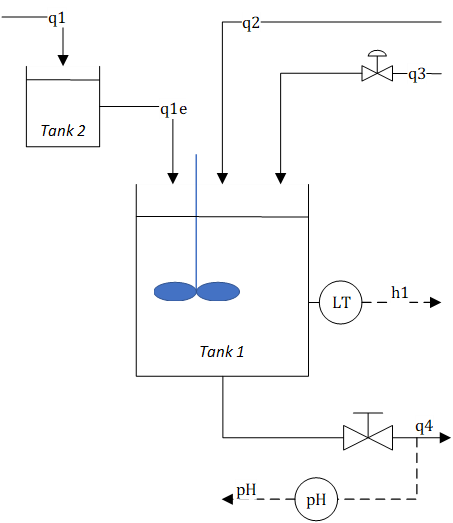}
		\caption{ $pH$ neutralization system scheme}
		\label{fig: sistemaph}
	\end{figure}

	The nonlinear system of study considered for our simulations is the $pH$ neutralization process represented in Fig. \ref{fig: sistemaph} where the goal is to keep the $pH$ of the solution at a neutral level ($pH=7$). 
	As it can be noticed, the system is composed of a principal tank, called reactor tank, in which the transformation takes place. Three flows enter the reactor:
	\begin{itemize}
		\item acid stream $q_{1e};$
		\item buffer flow $q_{2};$
		\item alkaline stream $q_{3}.$
	\end{itemize}
	where $q_1=q_{1e}$ is assumed to be constant since its dynamics is much faster than the main process one, the flows $q_1$ and $q_2$ are considered as unmeasured disturbances, while stream $q_{3}$ is the control input regulated by a valve. The output of the system is the stream $q_{4}$, that is the final solution from which the $pH$ is measured and controlled. The model of the process can be written in the state space form with $x \in \mathbb{R}^3$, $d \in \mathbb{R}$. For the full system equations we make reference to \cite{4790490}.
%	\begin{equation}
%	\begin{aligned}
%	x & \triangleq & [W_{a4}\quad W_{b4}\quad h] & \quad \text{states}\\[1ex]
%	d & \triangleq & q_{2} \qquad& \quad \text{disturbance}\\[1ex]
%	u_{a} & \triangleq & q_{3} \qquad& \quad \text{input}\\[1ex]
%	y & \triangleq & pH \qquad & \quad \text{output}\\[1ex]
%	\end{aligned}
%	\end{equation}
%	Thus, the process is described by the following nonlinear differential equations:
%	\begin{equation}
%		\begin{aligned}
%		\dot{x}(t)= f(x(t))+g(x(t)) & u_{a}(t)+p(x(t))d(t)	\\[1.5ex]
%		c(x(t),y(t)) & = 0	
%		\end{aligned}
%		\label{eq:processEq1}
%	\end{equation}
%	where $f,g$ and $p$ are functions of the state, $c$ is function of the state and of the output.
	\subsection{Identification}
	In order to identify the ESN model, the real plant has been excited over the whole operating conditions with a Multilevel Pseudo-Random Signal (MPRS) characterized by two different frequencies: one with switching period of $10$ seconds and the other one of $1000$ seconds.
	The choice of $10$ seconds is explained by the fact that, for the control of the neural network (ESN), the adopted sampling time is set to $10$ seconds. Moreover, we have to teach to the ESN how the real system behaves under "fast" changes of the input. On the other hand, the part of the input with period of $1000$ seconds, greater than the settling time, is used to well identify the steady state behaviour of the system.
	As for the amplitude, in the simulation for the $pH$ system we have used as the forcing input flowrate $q_3$ a MPRS signal with a minimum value of $12.7~ $ mL/s and a maximum one of $16.7~ $ mL/s. The corresponding $pH$ output of the process is in the interval $[6,8.65]$.
	\newline The input/output signals used to train the ESN have been acquired with a sampling time $T_s=10~s$ in order to have empirically from 20 to 40 samples in the part of the step response that goes from 0 to the settling time of the system. \newline
	Once the input/output sequences have been retrieved, an ESN with 300 states has been trained following the procedure explained in Section \ref{secTraining}. Thanks to the application of Algorithm 2, the final network with 188 states guarantees satisfactory performances despite the nearly $40$\% dimensionality reduction. A quantitative comparison of the trained networks' performances are reported in Table \ref{Table:fitting}, where the fitting value is computed as 
	\begin{equation}
	100 \left(1-\frac{\|Y_{sys}-Y\|_2}{\|Y_{sys}-mean(Y_{sys})\|_2} \right) \%
	\end{equation}
	
always with respect to validation data. We notice that if the training is performed only with steps 1-2 of Algorithm 2, the network shows negative performances, specifically the gain of the process is badly identified, and this motivates the adoption of the complete algorithm.

\begin{figure}[h]
		\centering
		\includegraphics[scale=0.42]{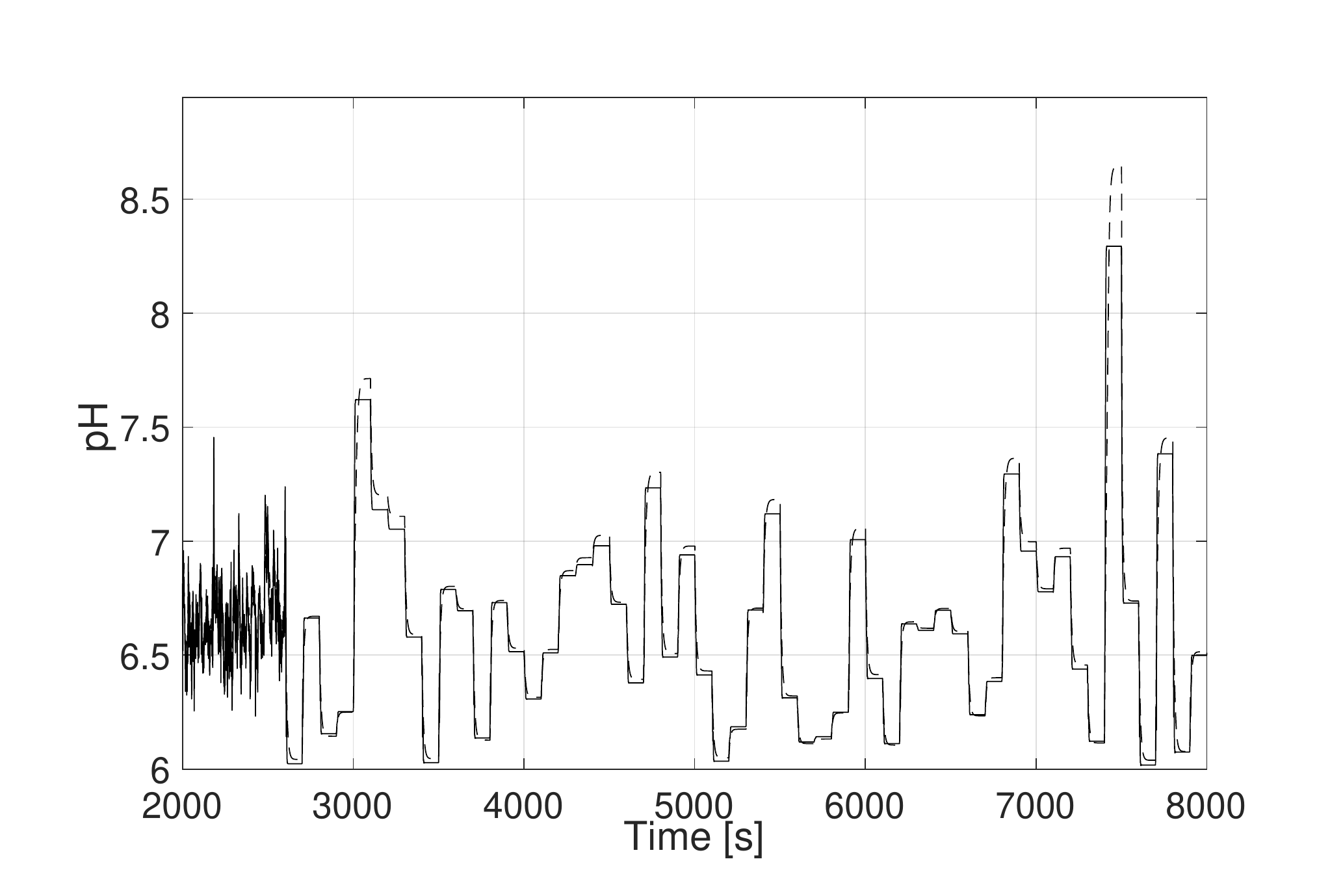}
		\caption{ Validation of the reduced network (188 states). dashed line: plant, solid line: identified model}
		\label{fig: identificazione}
	\end{figure}

		\begin{table}[h]
		\centering
		\renewcommand\arraystretch{1.2}
		\caption{\textit{pH system}:Comparison of the trained network performances} 
		\label{tab: parMPC} \small
		\begin{tabular}{ccc}
			\toprule
			\multicolumn{1}{c}{\textit{Algorithm}} &
			\multicolumn{1}{c}{\textit{Number of states $n$}} &
			\multicolumn{1}{c}{\textit{Fitting}}\\
			\midrule
			\textit{1} & 300 & 81.15\% \\
			\textit{2 (step \ref{LASSO})} & 300 & 71.75\%\\
			\textit{2 (step 1-2) } & 188 & -596\%\\			
			\textit{2 full} & 188 & 79.26\%\\			
			\bottomrule
			\label{Table:fitting}
		\end{tabular} 
	\end{table}	
\normalsize	
	As it is possible to see from Fig. \ref{fig: identificazione}, with the final identified ESN there is an error at steady state, that justifies also the introduction of the disturbance compensation in the predictive control algorithm discussed in Section \ref{mpc} and based on the ESN model.
	\subsection{Reference tracking}
\begin{figure}[h]
		\centering
		\includegraphics[scale=0.4]{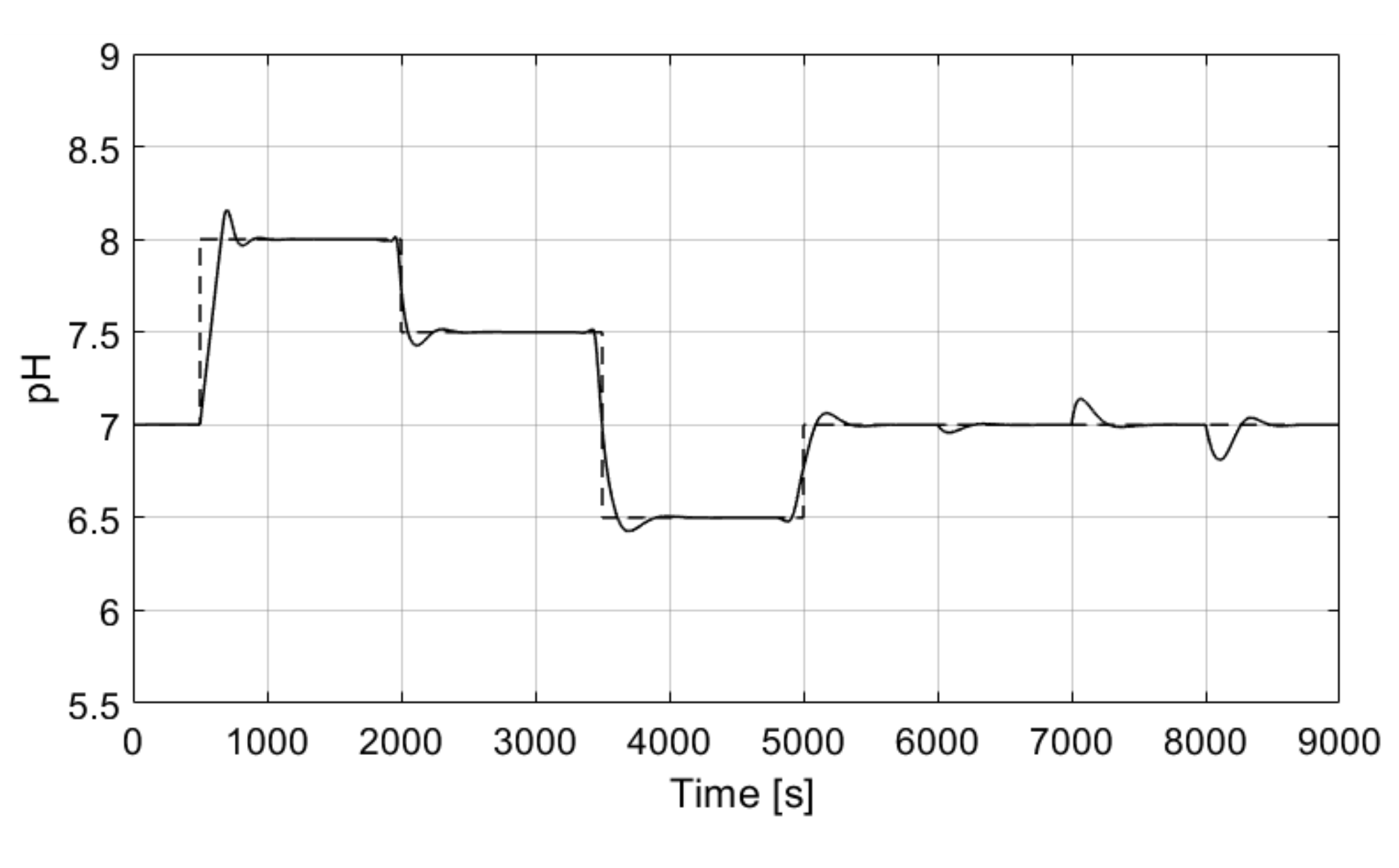}
		\caption{ $pH$ neutralization system scheme. dashed line: setpoint, solid line: output of the controlled system}
		\label{fig: controllo}
	\end{figure}

\begin{figure}[h]
		\centering
		\includegraphics[scale=0.42]{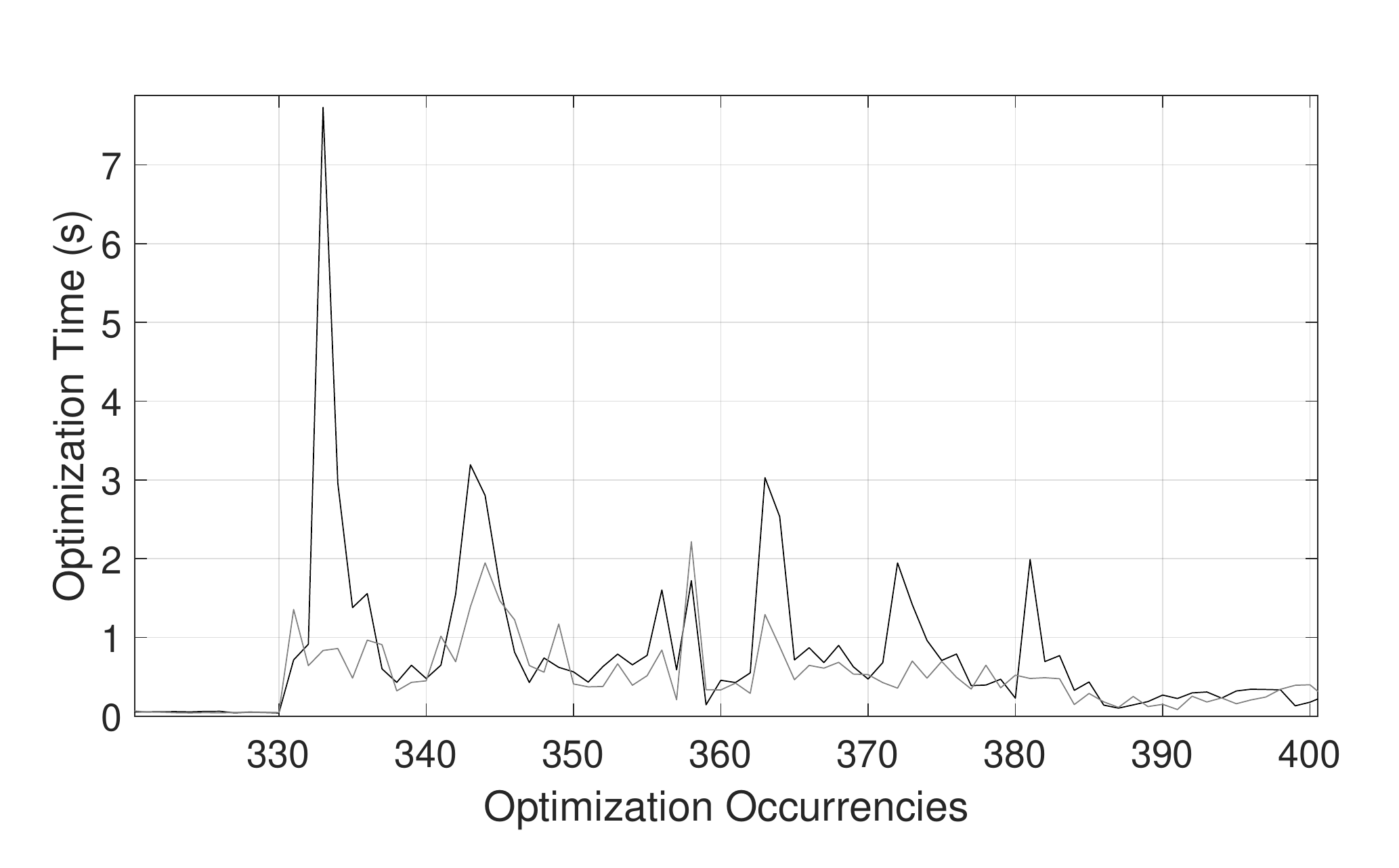}
		\caption{Comparison of times required to complete the nonlinear optimization. Black solid value: full ESN (300 states), grey line: reduced ESN (188 states) }
		\label{Fig:comparison_tempi}
	\end{figure}

	Once the identification of the model has been completed, the MPC described in Section \ref{mpc} has been used in order to control the original model of the pH process and, in particular, to track the changes of the reference signal $\bar{y}$ and to reject disturbances. 
	The parameters of the optimal control problem are listed for clarity in Table \ref{tab: parMPC}.
	\begin{table}[h]
		\centering
		\renewcommand\arraystretch{1.2}
		\caption{\textit{pH system}: Parameters for the MPC with model error compensation} 
		\label{tab: parMPC} \small
		\begin{tabular}{ccccc}
			\toprule
			\multicolumn{1}{c}{\textit{Parameter}} &
			\multicolumn{1}{c}{$N$} &
			\multicolumn{1}{c}{$T_s$} &
			\multicolumn{1}{c}{$Q$} &
			\multicolumn{1}{c}{$R$} \\
			\midrule
			\textit{Value} & $20$ & $10$ & $2$ & $1$ \\
			\bottomrule
		\end{tabular} 
	\end{table}	
	The predictive control performances are reported in Fig. \ref{fig: controllo} which shows the trend of the process output during reference tracking and in presence of step-wise non-modelled disturbance variations.
	The process has been first initialized with steady state input $\bar{u}=15.556~$ mL/s, then the output had to follow the steps in reference to values $[8\quad 7.5\quad 6.5\quad 7]$ at times $[500,2000,3500,5000]$ s and the non-modelled disturbance variations to values $[0.45\quad 0.85\quad 0.35]$ at instants $[6000,7000,8000]$.
	As it is possible to see, the output reaches quite fast the reference value with zero steady state error and with just small overshoots. Moreover, the control input counteracts the action of the non-modelled disturbance bringing back the system to its reference output value with zero offset.
	Then, we can conclude that the overall performances of the controlled process are satisfying.
Given the necessity to solve a nonlinear optimization program online, we checked the computational time required by MPC based on both the full network and the reduced one for the same control problem. The comparison is shown in Figure \ref{Fig:comparison_tempi}, and it shows that almost always the time required decreases for the network with a smaller number of states. The average value over the whole simulation is decreased from $0.35$ s to $0.25$ s, thus reporting nearly 30 \% reduction and motivating the dimensionality reduction of the network.

	\section{Conclusions}\label{Sec:Conclusion}
	In this paper we have focused our attention on Echo State networks, in order to test their applicability for control design purposes. First, a new result concerning stability-related properties of ESNs has been proven. Then, an algorithm for the reduction of the dimensionality and the training of ESNs has been proposed and exploited to obtain the identified model of a nonlinear case of study. Such model has been used for the design of an MPC regulator to tests the performances in reference tracking and in presence of a step-wise disturbance. Future work concerns the extension to MIMO systems and the derivation of additional theoretical properties.
	
	\section*{Appendix}
	In order to prove Theorem \ref{thm1} we first need to introduce some definitions.
	\begin{definition}[\cite{sohrab2003basic}]
		A real function $f:\mathbb{R}\to \mathbb{R}$ is called locally Lipschitz continuous if there exists a constant $L\geq 0$ such that, for any $x,y\in\mathbb{R}$ it holds:
		\begin{equation*}
		|f(x)-f(y)|\leq L|x-y|
		\end{equation*}
	\end{definition}
	\begin{remark}[\cite{sohrab2003basic}]
		An everywhere differentiable function $f:\mathbb{R}\to \mathbb{R}$ with ($L=\sup|f'(x)|$) is Lipschitz continuous if and only if it has bounded first derivative.
	\end{remark}
	\begin{definition}[\cite{sohrab2003basic}]
		If function $f$ is Lipschitz continuous on all the space $\mathbb{R}$, then $f$ is called globally Lipschitz continuous.
	\end{definition}
	\begin{remark}
		\label{lip}
		The function $f(x)=\tanh(x)$ is Lipschitz continuous since $\sup|f'(x)|=1$. Moreover it is also globally Lipschitz continuous with $L=1$.
	\end{remark}
	Now we can prove Theorem \ref{thm1}: 

\smallskip	

	\begin{proof}
		In order to prove $\delta GAS$ of system \eqref{net_state} we show the existence of a $\mathcal{KL}-Function$ in the initial states through a suitable $\delta GAS$ Lyapunov function. Let us consider $V(x(k),x'(k))=(x(k)-x'(k))^{T}(x(k)-x'(k))=\|x(k)-x'(k)\|_2^{2}$ as a candidate $\delta GAS$ Lyapunov function.
		We define $\omega(k)=(u(k),y(k))^T$ and $W_{in}=(W_u, W_y)$.
		From now on, for notational simplicity we drop the dependence on time $k$ which is implicit.
		\begin{equation*}
		\begin{split}
		&V(f(x,\omega),f(x',\omega))-V(x,x')=\dots\\
		&=\|f(x,\omega)-f(x',\omega)\|_2^{2}-\|x-x'\|_2^{2}
		\end{split}
		\end{equation*}
		By means of Lipschitz condition of the function \textit{tanh} we have:
		\begin{equation*}
		\begin{split}
		\|f&(x,\omega)-f(x',\omega)\|_2^{2}-\|x-x'\|_2^{2}\leq \dots\\
		\dots&\leq L^2\|W_{in}\omega+W_xx-W_{in}\omega-W_xx'\|_2^{2}-\|x-x'\|_2^{2}\\
		&\leq L^2\|W_x(x-x')\|_2^{2}-\|x-x'\|_2^{2}
		\end{split}
		\end{equation*}
		where $L=1$ is the Lipschitz constant of the function \textit{tanh}. Now, we can write:
		\begin{equation*}
		\begin{split}
		\|f(x,&\omega)-f(x',\omega)\|_2^{2}-\|x-x'\|_2^{2}\leq \dots\\
		\dots&\leq\|x-x'\|_{W_x^TW_x}^{2}-\|x-x'\|_2^{2}\\
		&\leq  \|x-x'\|^{2}_{W_x^{T}W_x-I}\\
		&\leq  \|x-x'\|^{2}_{-Q}
		\end{split}
		\end{equation*}
		where we introduced $Q=I-W_x^{T}W_x$.\newline
		Since by assumption $\|W_x\|^{2}<1$, all the eigenvalues of symmetric matrix $Q$ are $\lambda>0$. Therefore, $Q$ is a positive definite matrix, while $-Q$ is a negative definite one.
		
		Therefore
		\begin{equation*}
		V(f(x,\omega),f(x',\omega))-V(x,x')\leq-\|x-x'\|_{Q}^{2}
		\end{equation*}
		Then, with standard arguments we can write:
		\begin{equation*}
		\|x(k)-x'(k)\|_2\leq\alpha^k\|x(0)-x'(0)\|_2
		\end{equation*}
		where $\alpha=\sqrt{1-\lambda_{min}(Q)}$.\newline
		If we define $\beta(\|x(0)-x'(0)\|_2,k)=\alpha^k\|x(0)-x'(0)\|_2$ $\delta GAS$ is proved.
	\end{proof}
\smallskip
	\begin{property}[Alternative canonical form of ESN]\label{Prop_alternativeSS}
		In literature the mathematical model of the Echo State Networks is generally expressed by the following state and output equations, where $\xi(k+1)$ is the input vector:
		\begin{alignat}{2}
		x(k+1) & = \tanh(W_xx(k)+W_{u}\xi(k+1)+W_{y}y(k)) 
		\label{net_state1}\\[1.5ex]
		y(k) & = W_{out_{1}}x(k)+W_{out_{2}}\xi(k)
		\label{net_output1}
		\end{alignat}
		This formulation is equivalent to \eqref{net_state}-\eqref{net_output}, save for an additional state equation. In particular defining
		\begin{equation}
		\xi(k+1)=u(k)
\label{eq:auxiliaryeq}		
		\end{equation}
it follows that \eqref{net_state1} becomes \eqref{net_state}, and \eqref{net_output1} becomes \eqref{net_output}. \hfill $\square$
	\end{property}
	
\smallskip

\begin{property}\label{Prop_MN}
Mixed network are dynamical systems characterized by a state dynamic of the following form:
\begin{alignat}{2}
x_1(k+1)&=tanh(A_{11}x_1(k)+A_{12}x_2(k)+B_{1}u(k)) \label{MN1}\\
x_2(k+1)&=A_{21}x_1(k)+A_{22}x_2(k)+B_2u(k)\label{MN2}\\
y(k)&=C_1x_1(k)+C_2x_2(k)\label{MN3}
\end{alignat}

with $A_{11},A_{12},A_{21},A_{22},B_1,B_2,C_1,C_2$ matrices of suitable dimensions.
Comparing \eqref{net_state1},\eqref{eq:auxiliaryeq} and\eqref{net_output1} with \eqref{MN1},\eqref{MN2} and \eqref{MN3} respectively, the membership of ESN to Mixed Network is apparent.\hfill $\square$
\end{property}

\bibliography{Bibliografia}
\bibliographystyle{plain}

\end{document}